\documentclass{ws-procs975x65}

\begin{document}

\title{ON THE NO-GRAVITY LIMIT OF GRAVITY}

\author{JERZY KOWALSKI-GLIKMAN$^*$ and MICHA{\L} SZCZ\c{A}CHOR$^{**}$}

\address{Institute for Theoretical Physics ,\\
University of Wroclaw,\\
Wroclaw, Poland\\
$^*$E-mail: jkowalskiglikman@ift.uni.wroc.pl\\ $^{**}$E-mail:
misza@ift.uni.wroc.pl}

\begin{abstract}
We argue that Relative Locality may arise in the no gravity
$G\rightarrow0$ limit of gravity. In this limit gravity becomes a
topological field theory of the BF type that, after coupling to
particles, may effectively deform its dynamics. We briefly discuss
another  no gravity limit with a self dual ground state as well as
the topological ultra strong $G\rightarrow\infty$ one.
\end{abstract}

\keywords{Quantum Gravity Phenomenology; Relative Locality;
Topological field theories.}

\bodymatter\bigskip

It was rather clear from the very first days of the Quantum Gravity
Phenomenology research program \cite{AmelinoCamelia:1999zc} that
there might exist a class of potentially observable
 phenomena of the quantum gravitational origin
 exhibiting themselves in the form of minute deviations from the standard special
 relativistic kinematics and dynamics. This includes the models with Lorentz
 Invariance Violation  as well as the theories, in which the
 relativistic symmetries are still present, but become deformed.
 Both such models can be described in the framework of Relative
 Locality\cite{AmelinoCamelia:2011bm} and are generically characterized by
 the observation that the deviations from special relativity could be associated
 with the presence of a nontrivial geometry of momentum space.

 Quantum Gravity is a regime, in which both quantum and
 gravitational phenomena are `strong' i.e., the characteristic
 length scale of the process in question is of order of Planck
 length $\ell_P = \sqrt{\hbar\, G}$ and the characteristic energy
 scale is of order of Planck mass $M_P=\sqrt{\hbar/G}$. One can,
 however,
 consider a class of processes, in which the characteristic length scale is
 much larger than $\ell_P$ so that one can safely neglect the
 effects caused by spacetime foamy structure, but the energies are
 still close to Planckian.

 The presence of a scale is a prerequisite necessary for the emergence
 of a nontrivial geometry. In the case of theories with the scale
 of mass it is the momentum space that, in contrast with special
 relativity, may acquire a nontrivial geometry. This may result, in
 turn, in the emergence of new phenomena\cite{AmelinoCamelia:2011bm} that might be detectable in present
 experiments or in the foreseeable future.

 There are two ways one can get to the Relative Locality limit of
 Quantum Gravity. The first, more direct one, is to consider highly energetic
 processes with large characteristic length scales like the  ultra-Planckian scattering
 with the impact parameter much larger that $\ell_P$\cite{Verlinde:1991iu,
 Amati:1987uf}. Here we take the second, more indirect way,
 considering the $G\rightarrow0$, or no-gravity, limit of the classical
 general relativity (see\cite{Smolin:1992wj} for early discussion.)
 In this limit both gravitational and quantum effects are negligibly
 small and one could look for the Relative Locality regime, if
 present.

 The no-gravity
 limit can be most directly taken in the framework, in which gravity
 is constructed as a constrained topological BF theory\cite{Smolin:1998qp,
 Freidel:2005ak}. The starting point of this approach  is the BF topological field
 theory action with the (anti) de Sitter gauge group ($SO(3,2)$ or $SO(4,1)$, respectively) supplemented
 with the term that breaks this symmetry down to the local Lorentz
 one
 \begin{equation}\label{1}
    S =\frac1{16\pi}\, \int_{\cal M} B^{IJ} \wedge F_{IJ}(A) -
\frac{\beta}{2}\,B^{IJ} \wedge B_{IJ}- \frac{\alpha}4\,
\epsilon_{ijkl}\,
        B^{ij}\wedge B^{kl}
  \,,
 \end{equation}
where $I,J,\ldots$ are the (anti) de Sitter Lie algebra indices and
$i,j,\ldots$ are the ones of its Lorentz sub-algebra.

It should be mentioned that there is a natural coupling of gravity
defined by the action (\ref{1}) with point particles, which, as in
the case of 2+1 gravity, are represented by Wilson
lines\cite{Freidel:2006hv}.

Solving the algebraic field equations for the field $B^{IJ}$,
decomposing the connection $A^{IJ}$ into the Lorentz and
translational parts
\begin{equation}\label{2}
    A^{ij} = \omega^{ij}\,,\quad A^{i4} = \frac1{\ell_c}\, e^i
\end{equation}
and plugging the result into the action (\ref{1}) one gets the Holst
action of gravity with cosmological constant $\Lambda = 3/\ell_c^2$
appended by a number of topological terms (a linear combination of
Pontryagin, Euler, and Nieh-Yan classes)
\begin{equation}\label{3}
    32\pi G\, S=\int R^{ij}\wedge e^{k}\wedge e^{l}\,\epsilon_{ijkl}+\frac{\Lambda}{6}\int  e^{i}\wedge e^{j}\wedge e^{k}\wedge e^{l}\,
    \epsilon_{ijkl} +\frac{2}{\gamma}\int R^{ij}\wedge e_{i}\wedge
    e_{j}\,.
\end{equation}
The physical coupling constants, the Newton's constant $G$ and the
Immirzi-Barbero parameter $\gamma$ are related to the coupling
constants of the original action (\ref{1}) as follows
\begin{equation}\label{4}
    G = \frac{\alpha^2 + \beta^2}\alpha\, \frac1\Lambda\,,\quad
    \gamma=\frac\beta\alpha\,.
\end{equation}

There are two interesting limits of the theory described by
(\ref{1}) leading to the no-gravity regime. The first is  to fix an
arbitrary value of the Immirzi-Barbero parameter $\gamma\neq i$
(i.e., to assume the fixed relation $\beta=\gamma\, \alpha$) and
then to take the limit $\alpha\rightarrow0$. In this case we get
form (\ref{1}) a pure BF topological field theory, which is in many
respects similar to gravity in 2+1 dimensions.  It is well known
that in 2+1 dimensions gravity is topological. Moreover, after
coupling to particles and solving for (topological) degrees of
freedom of gravity one obtains, both
classically\cite{Matschull:1997du, Meusburger:2005mg} and quantum
mechanically\cite{Freidel:2005me} a deformed particle mechanics with
curved momentum space. It is hoped that the no-gravity limit of  the
physical 3+1 dimensional case  an analogous deformation of the
dynamics of particles coupled  gravity arises.

Another possibility is to take the Immirzi-Barbero parameter
$\gamma\rightarrow i$ which translates to $\beta \rightarrow \pm
i\alpha$ while keeping $\alpha$ arbitrary. In this case we have to
do with the (anti) self dual limit of the action (\ref{1}), which
reminds the setup discussed by Smolin\cite{Smolin:1992wj}. It would
be very interesting to analyze the perturbation theory (classical
and quantum) around this self dual ground state in powers of
$\delta\alpha = \alpha\pm i\beta$ given that, as stressed
in\cite{Freidel:2005ak}, such perturbation theory is going to be, by
construction, manifestly diffeomorphism invariant. However, it is
not clear if the self dual ground state, when coupled to particles
is going to exhibit any kind of the Relative Locality behavior.

Last, but not least one should mention the third possible limit that
can be read off from (\ref{4}). If we fix the value of $\beta$ and
then go with $\alpha$ to zero we obtain again a topological theory
(called sometime, misleadingly in the present context, the BF theory
with the cosmological constant $\beta$), which corresponds this time
to the ultra strong limit of gravity $G\rightarrow\infty$. There are
many indications (see\cite{Carlip:2009kf} for a recent review) that
in this limit spacetime becomes effectively lower dimensional. As
discussed in\cite{KowalskiGlikman:2008fj} there are good reasons to
believe that also in this limit gravity coupled to particles may
show signs of the Relative Locality behavior, but the full
understanding of this requires further studies.

 \section*{Acknowledgment}
 For JKG this work was supported in parts by the grant 2011/01/B/ST2/03354 and
by funds provided by the National Science Center under the agreement
DEC- 2011/02/A/ST2/00294; for MS this work was supported by the
grant 2011/01/N/ST2/00415.

\bibliographystyle{ws-procs975x65}

\begin{thebibliography}{99}

\bibitem{AmelinoCamelia:1999zc}
  G.~Amelino-Camelia,
  ``Are we at the dawn of quantum gravity phenomenology?,''
  Lect.\ Notes Phys.\  {\bf 541} (2000) 1
  [gr-qc/9910089].

\bibitem{AmelinoCamelia:2011bm}
  G.~Amelino-Camelia, L.~Freidel, J.~Kowalski-Glikman and L.~Smolin,
  Phys.\ Rev.\ D {\bf 84} (2011) 084010
  [arXiv:1101.0931 [hep-th]].

\bibitem{Verlinde:1991iu}
  H.~L.~Verlinde and E.~P.~Verlinde,
  ``Scattering at Planckian energies,''
  Nucl.\ Phys.\ B {\bf 371} (1992) 246
  [hep-th/9110017].

\bibitem{Amati:1987uf}
  D.~Amati, M.~Ciafaloni and G.~Veneziano,
  ``Classical and Quantum Gravity Effects from Planckian Energy Superstring Collisions,''
  Int.\ J.\ Mod.\ Phys.\ A {\bf 3} (1988) 1615.

\bibitem{Smolin:1992wj}
  L.~Smolin,
  ``The G(Newton) ---> 0 limit of Euclidean quantum gravity,''
  Class.\ Quant.\ Grav.\  {\bf 9} (1992) 883
  [hep-th/9202076].

\bibitem{Smolin:1998qp}
  L.~Smolin,
  ``A Holographic formulation of quantum general relativity,''
  Phys.\ Rev.\ D {\bf 61} (2000) 084007
  [hep-th/9808191].

\bibitem{Freidel:2005ak}
  L.~Freidel and A.~Starodubtsev,
  ``Quantum gravity in terms of topological observables,''
  hep-th/0501191.

\bibitem{Freidel:2006hv}
  L.~Freidel, J.~Kowalski-Glikman and A.~Starodubtsev,
  ``Particles as Wilson lines of gravitational field,''
  Phys.\ Rev.\ D {\bf 74} (2006) 084002
  [gr-qc/0607014].

\bibitem{Matschull:1997du}
  H.~J.~Matschull and M.~Welling,
  ``Quantum mechanics of a point particle in 2+1 dimensional gravity,''
  Class.\ Quant.\ Grav.\  {\bf 15}, 2981 (1998)
  [arXiv:gr-qc/9708054].

\bibitem{Meusburger:2005mg}
  C.~Meusburger and B.~J.~Schroers,
  ``Phase space structure of Chern-Simons theory with a non-standard puncture,''
  Nucl.\ Phys.\ B {\bf 738} (2006) 425
  [hep-th/0505143].

\bibitem{Freidel:2005me}
  L.~Freidel and E.~R.~Livine,
  ``Effective 3d quantum gravity and non-commutative quantum field theory,''
  Phys.\ Rev.\ Lett.\  {\bf 96}, 221301 (2006)
  [arXiv:hep-th/0512113].



\bibitem{Carlip:2009kf}
  S.~Carlip,
  ``Spontaneous Dimensional Reduction in Short-Distance Quantum Gravity?,''
  arXiv:0909.3329 [gr-qc].

\bibitem{KowalskiGlikman:2008fj}
  J.~Kowalski-Glikman and A.~Starodubtsev,
  ``Effective particle kinematics from Quantum Gravity,''
  Phys.\ Rev.\ D {\bf 78} (2008) 084039
  [arXiv:0808.2613 [gr-qc]].

\end{thebibliography}

\end{document}